# Predict And Prevent DDOS Attacks Using Machine Learning and Statistical Algorithms

Azadeh Golduzian[1]

[1]*Department of Mathematics and Statistics, University of New Mexico, NM 87106 USA*

*Corresponding author: Azadeh Golduzian ( agolduzian96@unm.edu).*

*This work was supported in part by the University of New Mexico, Albuquerque.*

A malicious attempt to exhaust a victim's resources to cause it to crash or halt its services is known as a distributed denial-of-service (DDoS) attack. DDOS attacks stop authorized users from accessing specific services available on the Internet. It targets varying components of a network layer and it is better to stop into layer 4 (transport layer) of the network before approaching a higher layer. This study uses several machine learning and statistical models to detect DDoS attacks from traces of traffic flow and suggests a method to prevent DDOS attacks. For this purpose, we used logistic regression, CNN, XGBoost, naive Bayes, AdaBoostClassifier, KNN, and random forest ML algorithms. In addition, data preprocessing was performed using three methods to identify the most relevant features. This paper explores the issue of improving the DDOS attack detection accuracy using the latest dataset named CICDDoS2019, which has over 50 million records. Because we employed an extensive dataset for this investigation, our findings are trustworthy and practical. Our target class (attack class) was imbalanced. Therefore, we used two techniques to deal with imbalanced data in machine learning. The XGboost machine learning model provided the best detection accuracy of (99.9999%) after applying the SMOTE approach to the target class, outperforming recently developed DDoS detection systems. To the best of our knowledge, no other research has worked on the most recent dataset with over 50 million records, addresses the statistical technique to select the most significant feature, has this high accuracy, and suggests ways to avoid DDOS attack

## I. INTRODUCTION

The phenomenon known as a "Distributed Denial-of-Service (DDoS) Attack" constitutes a formidable form of cybercrime that unleashes havoc by inundating a server with an overwhelming barrage of requests, effectively rendering online sites and services inaccessible to legitimate users. The insidious intent behind a DDoS attack lies in its capacity to disrupt the seamless provision of both internal and external services offered by a website [1]. What sets DDoS attacks apart in their potency is their utilization of a multitude of compromised computer systems as sources for the deluge of attack traffic. This assortment of exploited machines extends to encompass a gamut of devices, including computers and networked resources such as Internet of Things (IoT) devices [1]. The advent of the Internet of Things (IoT) heralds an interconnected realm wherein objects interlink to glean and exchange information autonomously, erasing the need for manual intervention [2]. However, as the IoT's omnipresence burgeons, the proliferation of remote employees and the surge in Internet-connected devices come with a caveat. IoT devices, while prolific, may not consistently uphold robust security measures, leaving the networks they permeate susceptible to infiltration and malicious attacks.

Hence, the imperativeness of DDoS prediction and fortification looms large. An intriguing dichotomy emerges when inspecting DDoS attack trends; the frequency of such attacks witnessed a discernible dip from the onset of 2021 until its culmination, while maintaining relative consistency over the preceding biennium. Remarkably, 2021 exhibited a mere 3% reduction compared to the prior year. Paradoxically, as the attack frequency wanes, the magnitude of these attacks experiences exponential growth [3]. Envision a malevolent hacker striving to incapacitate a service. Here, the hacker's modus operandi veers from employing a solitary computer and static IP address; instead, an arsenal of diverse computers, each equipped with distinct IP addresses, is wielded to elude security measures. A pivotal hallmark of DDOS attacks is their orchestrated orchestration from a multitude of hosts, encompassing even the manipulation of your server to assail another. Often orchestrated by botnets, these attacks unfold through networks of automated robots or computers, each programmed to execute specific tasks—a landscape where the term "zombies" finds relevance. Historically, the roots of DDoS trace back to 1998, though the full impact remained obscure until July 1999, when influential organizations and corporate entities endured the brunt of these assaults [4]. The repercussions of such attacks are far-reaching, with organizations and communications infrastructures susceptible to debilitating disruptions that may extend to minutes or even hours, if proactive protection mechanisms are not in place. Thus, the urgency to fortify digital bastions against these evolving threats has never been more critical.

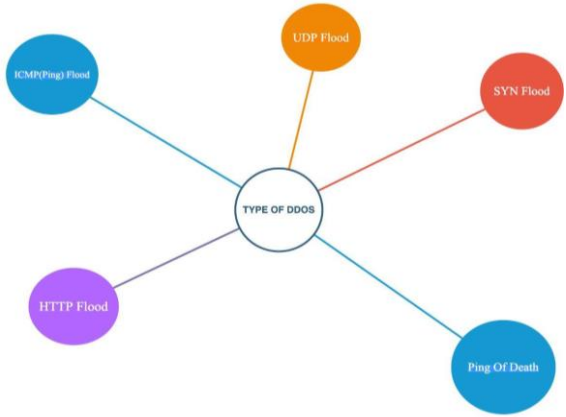

FIGURE 1. Types of DDOS attacks

In addition, there are capacity issues for businesses that provide defense systems to stop this attack. Fig. 1. presented some DDoS attacks.

1- UDP Flood 2- ICMP(Ping) Flood 3- SYN Flood 4- Ping Of Death 5- HTTP Flood. I will give a brief description of each attack:

- **UDP Flood**: In this attack, the attacker sets up random ports on the target network by sending IP packets containing UDP datagrams. The victim system repeatedly attempts to prevent the UDP packet from responding in an attempt to match each datagram technique to a program, but it is unable to do so and will eventually wear out and fail.
- **HTTP Flood**: This attack uses several, seemingly genuine HTTP GET or POST requests to target an application or web server. These inquiries are frequently made to help criminals avoid being discovered by learning crucial details about their intended victims before an attack.
- **Ping Flood and Ping of Death**: Another typical flood attack exploits many ICMP echo queries. The target system attempts to react to numerous requests, eventually restricting its network bandwidth because each ping given requires a cross-response that constitutes the same number of packets to be returned. Another variation of this attack known as "ping of death" causes the operating system to crash by having the victim send ping packets with the incorrect format and shape.
- **SYN Flood**: Three-way communication between two systems is necessary for every TCP session. Using an SYN flood, the attacker rapidly overwhelms the victim with connection requests, so numerous that it can no longer handle them, causing network saturation. This occurs when the host sends a large number of TCP/SYN packets with a forged sender address. Each of these packets functions as a connection request, causing the server to maintain several open half connections. The server sends or returns TCP/SYN-ACK packets to wait for response packets from the sender's address; however, no response is returned because the sender's address is fake.

This paper aims to find the best algorithms to detect DDOS attacks, separate them from regular traffic, and identify the features that are most relevant to see so we can prevent them in advance. The significant contributions of this study are as follows.

- Analysis of the latest dataset named CICDDoS2019 with over 50 million records and 88 features
- Use statistics and machine learning algorithms to determine the best relevant features(feature selection)
- To improve the effectiveness of DDoS detection, various machine learning models were used during the training process.
- Test the machine-learning models and use the model with the best accuracy score among other methods.

## II. Related works

The closest competitor to this study and any similar models that used the CICDDoS2019 dataset are briefly described in this section. Research related to [5] surveys recently created machine-learning-based DDoS detection methods. The authors of [6, 7, 8] suggested the naive Bayes model as the DDoS detection method. In contrast, this work [9, 10, 11] applied a support vector machine model to identify the presence of DDoS attacks.

In addition, as shown in [12, 13], the decision tree algorithm has been used to detect DDoS attacks. The authors of [14] used a deep neural network (DNN) as a deep learning technique to identify DDoS attacks in a sample of packets from network traffic. Because the DNN model includes feature extraction and classification techniques, it can operate quickly and with a high degree of detection accuracy even with tiny samples. The CICDDoS2019 dataset, which contains multiple DDoS attack types developed in 2019, was used by the authors to conduct the tests. The proposed system achieved an accuracy rate of 94.57 percent using a deep learning model. Zeeshan Ahmad et al.

[15] proposed a scientific classification approach that depends on well-known ML and DL processes in the planning network-based IDS (NIDS) framework. To the best of our knowledge, no other research has worked on the most recent dataset with over 50 million records, addresses the statistical technique to select the most significant feature, has this high accuracy, suggests ways to avoid DDOS attacks, and tackles these issues.

III. PROPOSED MODEL

This study aims to create an accurate DDoS attack detection algorithm with a low false-positive rate. Here, a model based on a collection of seven classifiers for machine learning is provided. The chosen classifiers were naive Bayes, KNN, logistic regression, CNN, XGboost, AdaBoost, and random forest. All seven algorithms in our model operate independently and produce a unique data model. The outputs of the seven classifiers were combined using accuracy, precision, and recall techniques to arrive at the model's outcome. To train the model, the CIC-DDoS2019 dataset was used. This dataset contained 88 features. The feature set of the training dataset should be condensed, which is accomplished using the feature selection algorithm, "ANOVA, ExtraTreeClassifier, and logistic regression."

A. *Methodology*

In the context of this study, we employed the CICDDoS2019 dataset as the foundation for our DDoS attack detection endeavors. The systematic process employed for DDoS detection is intricately portrayed in Figure 2, offering a visual representation of our investigative methodology. Initiating this process, the foremost step involved meticulous examination of a sufficiently comprehensive DDoS dataset. Specifically, all 11 CSV files were meticulously amalgamated to create a unified and holistic dataset that would serve as the bedrock of our analysis. Subsequent to dataset compilation, our focus turned to refining and curating the data. A critical aspect of this stage entailed the selection of essential characteristics and features that would play a pivotal role in our analysis. This was accomplished through the application of three distinct feature-extraction techniques, each tailored to extract key attributes that would contribute to the precision of our DDoS detection framework. The culmination of these preparatory steps led to the development of a robust machine learning system, meticulously designed to detect DDoS attacks within the dataset. To gauge the efficacy of this system, we subjected it to comprehensive testing using an array of machine learning models. The testing phase encompassed both training and validation components, collectively serving to quantify the accuracy and reliability of the model's performance. This multistep approach, spanning dataset curation, feature extraction, and model training and testing, constitutes a comprehensive framework designed to empower accurate DDoS attack detection. By iteratively refining our methods and harnessing the potential of machine learning, we endeavor to fortify the efficacy of our approach and contribute to the domain of cybersecurity research.

B. *CICDDoS2019 Dataset*

CICDDoS2019 represents the most up-to-date version of the dataset currently accessible for analysis. Comprising a comprehensive collection of 88 distinct features and an extensive dataset encompassing over 50 million records, this repository includes data on both benign and denial-of-service (DoS) flows. The dataset's uniqueness lies in its incorporation of network traffic analysis outcomes, which have been meticulously annotated based on a variety of attributes such as source ports, inbound and outbound IP addresses, destination ports, and protocols. A focal point of our investigation centers on the attribute denoted as "Label." This particular attribute serves as our target of interest, dichotomizing data into two distinct classes: $Y=1$, denoting the attack class, and $Y=0$, characterizing the benign class. Notably, the observations attributed to the attack class surpass those of the benign class, thus creating an inherent imbalance within the dataset. To address the challenge posed by imbalanced classes, a meticulous approach was taken. Specifically, a combination of oversampling and undersampling techniques was deployed. Among these methodologies, the Synthetic Minority Over-sampling Technique (SMOTE) emerged as the most effective strategy for our data. By employing SMOTE, a balanced representation was achieved, resulting in approximately 50% of the data constituting attack instances, while the remaining 50% corresponded to benign instances. This strategic application of oversampling has been visualized and is illustratively depicted in Figure 3, underscoring the equilibrium achieved within the dataset after employing the SMOTE technique. This step not only addresses the imbalanced nature of the dataset but also lays the foundation for more robust and accurate analytical outcomes.

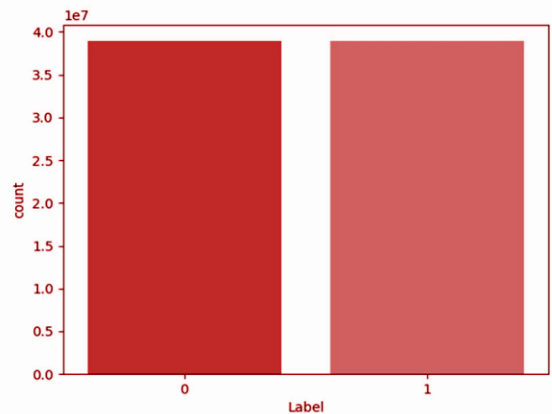

FIGURE 2. After Over Sampling

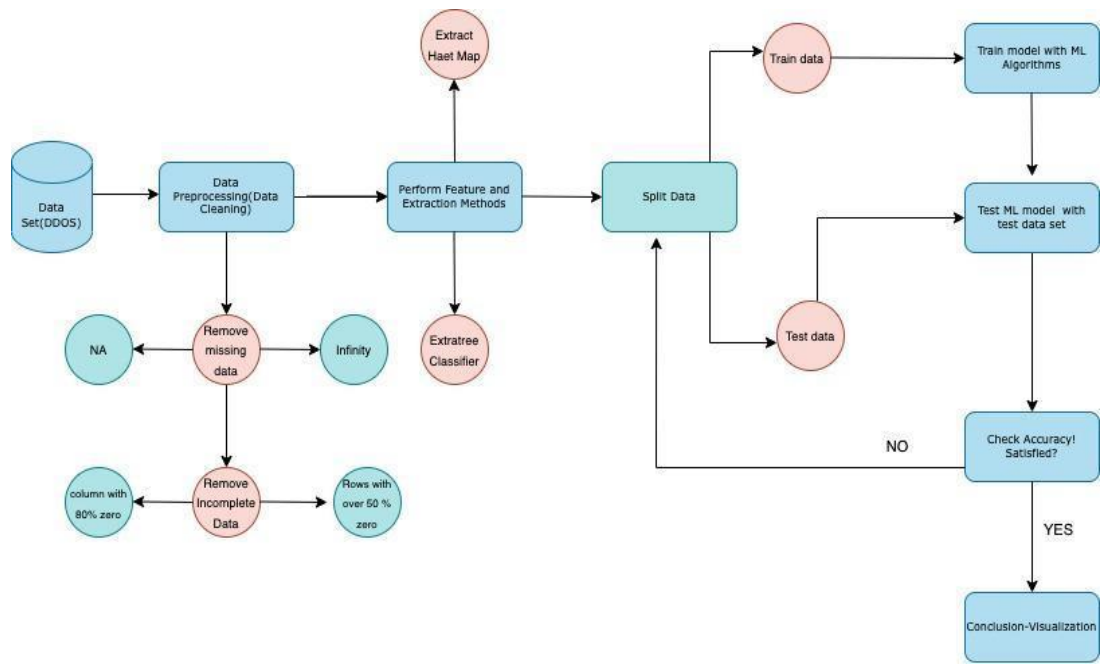

FIGURE 3. ML Data Process

There are 11 CSV files in the CICDDoS2019 dataset and 11 attacks, namely, **TDTP, Syn, DrDoS_UDP, DrDoS_DNS, DrDoS_LDAP, DrDoS_SSDP, DrDoS_MSSQL, DrDoS_NetBIOS, UDP-lag, DrDoS_SSDP, and DrDoS_NTP.**

To perform a machine-learning model, I split the data into training and test datasets, first training the model with the training dataset, and then testing our model. Testing the model is key to understanding the performance of the suggested model.

### C. Data preprocessing

A data-mining technique called data preparation, which is both practical and effective, is used to format the raw data. The steps involved in data Preprocessing were **1-Data cleaning, 2-Data transformation, and 3- data reduction**.

Most machine learning models only work with numeric values.

In this step, we converted non-numeric values into numbers. Feature set selection is performed using a heatmap matrix, tree classifier, and an additional logistic approach. Should reduce the features used to train the model.

Using more features can lead to low accuracy, whereas using fewer features can lead to a high false-positive rate. It should balance the number of features to obtain a model with high accuracy but a low false-positive rate. In the dataset, attributes (columns) that contained mostly zero values were removed because they negatively affected the models.

Furthermore, in our pursuit of refining data quality, it was imperative to address the presence of rows harboring either missing values or infinite values. Recognizing that certain algorithms are ill-equipped to handle such anomalies, and that these aberrations can potentially impede machine learning efficiencies, we consciensiously omitted these rows from our analysis.

As emphasized earlier, the expanse of the CICDDoS2019 dataset unfurls across a staggering 88 distinct features. The sheer magnitude of these features ushers in complexities during both training and predictive phases. Navigating this intricate landscape, therefore, necessitates a judicious curation of features—choosing the ones most germane to the model's objectives. Opting to engage with pertinent features as opposed to encompassing all features, irrespective of their relevance, furnishes several invaluable advantages. Notably, this selective approach enhances the velocity of model training, allocates more temporal resources for the predictive task, amplifies prediction accuracy, and decisively mitigates the specter of overfitting.

The ensuing section assumes the mantle of elucidating the strategies employed to distill indispensable features from the labyrinthine CICDDoS2019 dataset.

A panoramic view of these strategies is encapsulated within Figure 4, delineating a quartet of categories: Filter Methods: Delineating techniques that derive feature relevance independent of the chosen machine learning model.

Wrapper Methods: Encompassing methodologies that evaluate feature subsets based on model performance, often employing a predictive model as a yardstick. Embedded Methods: Entailing methods where feature selection transpires as an inherent part of the model-building process. Hybrid Methods: Navigating a fusion of the aforementioned strategies to harness their collective strengths. In a bid to orchestrate this feature curation symphony, our approach gravitated towards a composite tapestry of techniques. Filtering, embedded, and elimination features were harnessed in tandem, buttressed by the bedrock of logistic regression—a formidable statistical methodology renowned for its robust analytical capabilities. The subsequent phases of our analysis were artfully choreographed to not only isolate essential features but to also synergize these features in a manner that underpins the veracity and robustness of our predictive model.

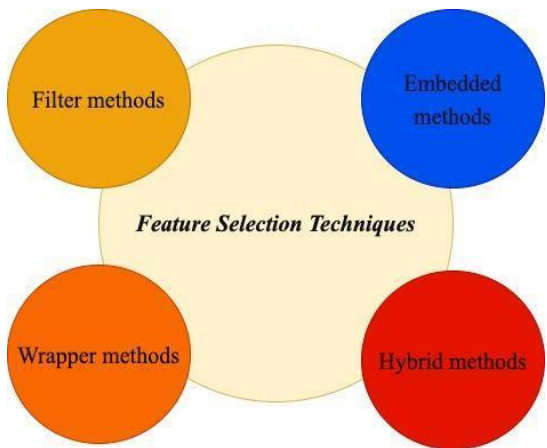

FIGURE 4. Feature Selection

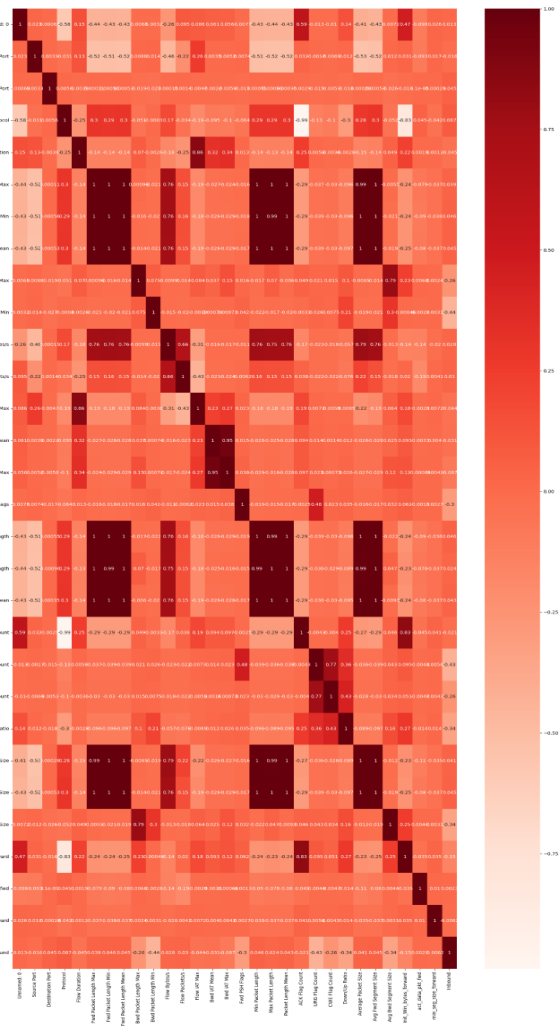

FIGURE 5. Heat Map

### 1. Filtering

In contrast to emphasizing the cross-validation performance, filter approaches delve into the inherent traits of features, gauged through univariate statistics. Unlike wrapper methods, these techniques function with swiftness and computational efficiency, making them a pragmatic choice. Particularly when dealing with data with a high number of dimensions, filter methods prove to be more economically viable in terms of computation. Figure 5 unveils the elegance of the heat map matrix, a quintessential embodiment of the filter approach. Within this matrix, the interplay of the 30 features is vividly portrayed, painted with shades of correlation.

This correlation—a measure of linear connection between multiple variables—can unravel the ability of one variable to forecast another. This is akin to deciphering the symphony within data, where one note holds predictive potential for another. The tapestry of desirable features is woven with threads of correlation that intricately connect them to the target. A strong correlation between a variable and the target makes it an attractive candidate for selection.

However, this isn't a simple weave; the variables should harmonize through uncorrelation amongst themselves, while maintaining a meaningful correlation with the ultimate goal. Just like skilled musicians in an orchestra, variables must strike a balance—each contributing in a harmonious fashion to create the symphony of predictive excellence.

### 2. Embedded

The methods we used in this study cleverly combined the strengths of two approaches—wrapper and filter—by considering how different features work together while also keeping the calculations manageable. Embedded systems, which work step by step in training our model, carefully picked out the most helpful parts at each stage. We used a special tool called a decision tree classifier to figure out which features were the most important. In a big chart (Figure 6), we highlighted the top 30 features that mattered the most out of the huge dataset of 50 million records. When it comes to picking the right features, we didn't just guess. We tried different combinations of 30, 20, and 10 best features to see what gave us the best results. Figure 6 also showed us that six features— "Inbound," "URG Flag Count," "Destination Port," "Avg Bwd Segment Size," "min-seg-size-forward," and "Source Port"— stood out as the most useful. But not all features are created equal. As shown in Figure 6, only 10 out of the 30 features turned out to be really good at helping us make predictions. The rest weren't as useful for our goal. This discovery helped us focus on what really matters in making accurate predictions.

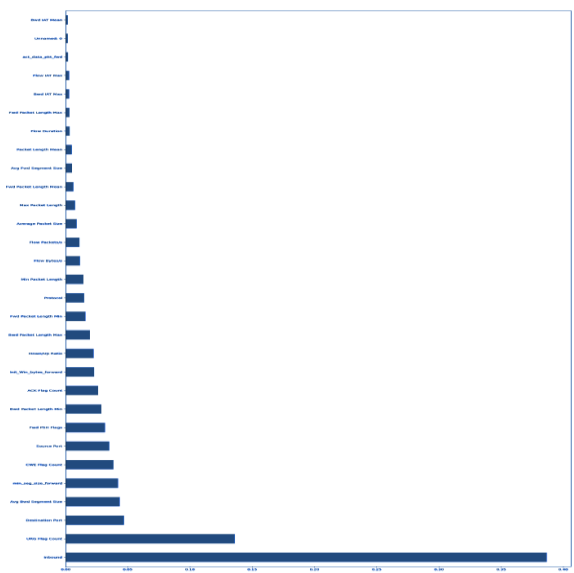

FIGURE 6. The best 30 relevant features

### 3. Feature elimination with logistic regression

As shown in fig 7, the p-values for most of the variables are smaller than 0.05, except for four variables. Therefore, when we wish to test our model with 20 and 10 features, these are the candidate characteristics that we will eliminate. Here, I must point out that we cannot make a decision based only on the logistic regression; rather, we must also take into account the heat map and the extra tree classifier because statisticians believe that p-values that are too low (0.0000) and too high (1.000) are questionable. Again, we see that "Inbound" has the highest effect and "Fwd packet length means" is the one that needs to be removed from our model.

### D. DDOS ML Model

Using individual classifiers, the model classifies the data individually. These classifiers operate in parallel and generate multiple models from a training dataset. Although many other classifiers are available for machine learning, we used seven classifiers.

#### 1. Naive Bayes

Indeed, the probabilistic learning model known as Naive Bayes (NB) finds its niche in the realm of machine learning, where its prowess lies in data classification. By following a sequence of calculations rooted in probability principles, Naive Bayes takes in data and efficiently sorts it into predefined categories. The underlying foundation is Bayes' theorem, a pivotal theorem in probability theory. One of the charming traits of Naive Bayes is its simplicity and speed of implementation. However, this efficiency comes with a prerequisite—it assumes that the predictors are independent of each other, which is a bit of an idealized assumption. So, how does this classification wonder work? It starts with feeding a chunk of training data into the system. This training data is a compilation of samples, each tagged with a specific class. The essence of NB lies in learning from these examples. When it comes to testing, the system processes the test data using methods learned from the training data. It's like following a well-practiced recipe—each ingredient (feature) contributes its flavor to the final dish (class prediction). The more elaborate your training dataset, the more adept the system becomes at dishing out accurate predictions for the test data's class. It's as if the system becomes a seasoned chef, confidently picking out the main ingredients from a jumble of flavors. So, in this culinary analogy of machine learning, Naive Bayes serves as the master chef—quick, efficient, and reliable, as long as the ingredients play along with its independence assumption.

```
                        Coef.      Std.Err.        z      P>|z|      [0.025           0.975]
------------------------------------------------------------------------------------------------
Unnamed: 0              0.0000      0.0000    495.0444   0.0000      0.0000           0.0000
Source Port            -0.0000      0.0000    -83.9446   0.0000     -0.0000          -0.0000
Destination Port       -0.0000      0.0000   -310.2489   0.0000     -0.0000          -0.0000
Protocol               -1.0246      0.0009  -1095.7349   0.0000     -1.0264          -1.0228
Flow Duration          -0.0000      0.0000   -148.9616   0.0000     -0.0000          -0.0000
Fwd Packet Length Max  -0.0299      0.0002   -177.1541   0.0000     -0.0303          -0.0296
Fwd Packet Length Min  -0.0820         nan        nan      nan         nan             nan
Fwd Packet Length Mean  0.0398 4086226714.7359  0.0000  1.0000  -8008857193.5079  8008857193.5876
Flow Bytes/s            0.0000      0.0000    221.3459   0.0000      0.0000           0.0000
Flow Packets/s         -0.0000      0.0000   -153.0026   0.0000     -0.0000          -0.0000
Flow IAT Max           -0.0000      0.0000   -202.4574   0.0000     -0.0000          -0.0000
Bwd IAT Mean            0.0000      0.0000    121.9669   0.0000      0.0000           0.0000
Bwd IAT Max            -0.0000      0.0000   -123.8993   0.0000     -0.0000          -0.0000
Fwd PSH Flags         -65.0551      0.1765   -368.5805   0.0000    -65.4011         -64.7092
Min Packet Length       0.0879         nan        nan      nan         nan             nan
Max Packet Length      -0.0139      0.0002    -64.3343   0.0000     -0.0144          -0.0135
Packet Length Mean     -0.0396      0.0003   -154.6618   0.0000     -0.0401          -0.0391
Packet Length Mean     -0.0396      0.0003   -154.6618   0.0000     -0.0401          -0.0391
ACK Flag Count          4.1954      0.0055    763.9291   0.0000      4.1846           4.2062
Average Packet Size     0.0539      0.0002    310.4721   0.0000      0.0535           0.0542
Avg Fwd Segment Size    0.0298 4086226714.7361  0.0000  1.0000  -8008857193.5185  8008857193.5780
Avg Bwd Segment Size    0.0058      0.0003     19.0479   0.0000      0.0052           0.0063
act_data_pkt_fwd        0.2907      0.0013    223.2098   0.0000      0.2882           0.2933
min_seg_size_forward    0.0000      0.0000     63.7411   0.0000      0.0000           0.0000
Inbound                 4.6551      0.0066    703.3782   0.0000      4.6422           4.6681
```

FIGURE 7. P-value of Selected Features

### 2. KNN

KNN is the abbreviation for k-nearest neighbors. It is a nonparametric algorithm based on a supervised learning technique that can be used to solve classification and regression problems. It stores all the existing data and classifies a new data point based on its similarity. When new data are introduced, it determines the class of the new data by looking at its K-nearest neighbors. The Manhattan, Minkowski, and Euclidean distance functions are used to determine the distance between two data points. The Euclidean distance function was used in this study. Similarity between the samples to be classified and the samples in the classes were detected. When faced with new data, the distance of this data to the training set data was calculated separately using the Euclidean function. The classification set was then created by selecting k datasets from the smallest distance. The number of KNN (k) neighbors is based on the classification value.

### 3. Random Forest

It consists of several decision trees that are independently trained on a random subset of labeled data. Random forest works well because many relatively separate trees perform better than any individual model.

### 4. XGBOOST

let's delve further into the world of the XGBoost learning model, a powerhouse that operates on the foundation of trees but brings a turbocharged performance that's almost too good to believe. In fact, its speed is so impressive that it can be up to 100 times faster than other models in its league. Imagine a racing car on a machine learning track—XGBoost would be the one leaving all others in the dust. The allure of XGBoost lies not just in its breakneck pace, but in its harmonious blend of attributes. It's not just a speed demon; it's also scalable, accommodating a vast volume of data without breaking a sweat. Efficiency dances hand in hand with simplicity, making it accessible to a wide range of users, from newcomers to seasoned data scientists. One of the remarkable traits of XGBoost is its reliability when it comes to handling large datasets. It's like having a reliable workhorse that can manage massive loads with ease, ensuring that data analysis doesn't buckle under the weight of sheer volume. When you're navigating through an ocean of data, XGBoost is the trusty ship that keeps you sailing smoothly. At the core of its operation lies probability—a crucial element in its decision-making process.

Each step XGBoost takes, each branching decision it makes, is rooted in probability. It's like a well-tailored suit, where each stitch is carefully calculated to bring out the best fit—except in this case, it's the best fit for predicting outcomes with an astute understanding of likelihood. So, in the realm of machine learning, where models are akin to skilled artisans crafting masterpieces, XGBoost stands out as the sprinter among the runners, the reliable partner amidst the data deluge, and the probability-savvy maestro composing predictions with precision [15].

### 5. Logistic Regression (Statistical Method)

The probability of a response variable was predicted using a supervised learning classification algorithm known as logistic regression. Because the nature of the dependent variable is binary, only two possible classes exist. Here, if the traffic flow is considered an attack, its value is 0. In other words, P(Y=1) was predicted by a regression model as a function of X.

### 6. Convolutional Neural Network (CNN)

A convolutional neural network (CNN/ConvNet) is a deep learning technique that takes an input image and assigns importance (learnable weights and bias) to each of the objects/aspects in the image while also distinguishing which ones are related to one another. Comparatively speaking, the ConvNet algorithm requires less "pre-processing" than other classification techniques. ConvNets can learn these filters/features with sufficient practice; however, the filters of the original approaches are manually engineered. The main advantage of CNN over its predecessors is that it automatically detects significant features without human supervision, making it the most used [16].

### 7. AdaBoost

AdaBoost is an ensemble learning technique (Statistical Classification) that was initially developed to boost the performance of binary classifiers (sometimes referred to as "meta-learning"). AdaBoost uses an iterative process to improve the poor classifiers by learning from their errors.

## IV. RESULTS AND DISCUSSION

In this section, we summarize the findings from various tests conducted to evaluate the effectiveness of various machine learning models. Moreover, it introduces an analysis to determine the differences between ML models.

### A. *Performance Evaluation*

To evaluate the effectiveness of the deployed DDoS detection system, several factors were considered, including

**Confusion matrix:** A confusion matrix table is used to describe the performance of a classification system. The effectiveness of the classification was represented and summarized using a confusion matrix.
- TP: You predicted positive, and it's true
- FP(Type 1 Error): You predicted positive, and it's false
- TN: You predicted negative, and it's true
- FN(Type 2 Error): You predicted negative, and it's false

**Accuracy**: Evaluation of a model's performance in a dataset to find relationships and patterns based on input data, also known as training data.

**Recall:** Recall is determined as the proportion of positive samples correctly identified as positive for all positive samples.

$$Recall = \frac{TP}{TP + FN}$$

**F1-Score**: The harmonic mean of recall and precision is known as the F1-score. The following formula combines the recall and precision into a single formula:

$$F1\text{-}score = 2 \times \frac{Precision \times Recall}{Precision + Recall}$$

**Precision**: Metrics such as precision and recall allow us to assess how well a classification model predicts outcomes for a given class of interest, or "positive class.". While recall measures the degree of the error caused by false negatives (FNs), precision measures the degree of the error caused by False Positives (FPs).

$$Precision = \frac{TP}{TP + FP}$$

## B. *Evaluation of ML models*

As mentioned previously in the preceding part, we evaluated seven different machine learning models, and each model was assessed using 30, 20, and ten features. Among models with 30 elements, XGBoost provided the best accuracy (99.99996%). With 20 features in the Rf model, it offers the best accuracy of (99.99999%) with a precision of 1, and KNN with an accuracy of (99.98%). With ten features, XGboost and RF had the same accuracy of (99.99%). The CNN models with 30 features produced an accuracy of (84.75%). XGboost and RF were the best ML models in our study. The detection accuracies for the top six machine-learning models using 30 characteristics are displayed in Table 1. We can save time and money by reducing the number of features as much as possible, for example, to only five. As shown in Fig. 8, if the firewall can identify attacks with only five features with over 80 % rate, we can buy some time and prevent the server from going down. It takes a long time, and it is not reasonable for the firewall to check 30 or 20 features to determine if they are attacks or benign.

Next, we discuss the precision, which was evaluated for each machine-learning model. Precision is the frequency at which the machine-learning model can predict the correct response. The XGBoost model with a 30-feature set provided the highest level of precision (100% ). However, the RF model obtained the highest precision result with a 20-feature group (100%), whereas the KNN model achieved the best precision result with a 20-feature set (99.99%) and the CNN model provided the best precision result with a 20-feature set (98.99 %). Consequently, the Random Forest model with a 20-feature set had the best precision results.

| ML Algorithms with 30 Feature | Accuracy | Precision | Recall | FN |
|---|---|---|---|---|
| XGBoost | 99.99996 | 1.00 | 1.00 | 3 |
| AdaBoost | 99.97 | 99.99 | 99.97 | 1227 |
| KNN | 99.73 | 99.97 | 99.75 | 119190 |
| Logistic Regression | 80.63 | 99.97 | 80.63 | 1172949 |
| RF | 99.9998 | 1.00 | 99.99 | 4 |
| Naive Bayes | 90.08 | 99.96 | 90.11 | 598920 |

Table. 1 Results of 6 ML Algorithms

## C. *Visual Explanation*

I will begin by using an approach known as individual conditional expectation (ICE) plots. They are straightforward to use and demonstrate how the forecast varies as the feature values change. They are comparable to partial dependence graphs, but as ICE plots show one line per instance, they go one step further and illustrate the heterogeneous effects. figure 9 shows that there is a positive relationship between the 'Inbound' and our target. The thick red line is the Partial dependency plot, which shows the change in the average prediction as we vary the "Inbound" feature. Inbound is the first and most significant feature that has a beneficial effect on attacks. This demonstrates that whenever attacks occur, they are highly predictable as attacks if Inbound <= 0.5 (Fig. 10); however, to be more accurate, we need to build a decision tree. To simplify, I train an exemplary decision tree and make predictions based on our random forest regressor. This tree is based on 30 million data, and we see that the first split is at the feature 'Inbound,' followed by the 'Source port' and 'Destination port. 'If you recall, these were the three most essential features picked by the random forest, heat map, and extra tree classifier. In addition, the p-value table calculated zero values for these three features.

## V. CONCLUSION AND FUTURE WORK

Within the landscape of network security, DDoS attacks have emerged as a prominent adversary, especially within expansive networks. Astonishingly, a staggering 64 percent of these attacks have been found to impact the operational security of the very servers that are fundamental to the functioning of our most widely used systems [17]. The ubiquity of DDoS attacks in large networks has propelled us on a mission to harness the power of machine learning to predict and subsequently thwart these malevolent activities. By delving into the intricacies of the ML model, we've been able to gain predictive insights into potential DDoS attacks. Armed with the knowledge of pertinent features, we're embarking on a journey to not just predict but actively prevent these attacks from wreaking havoc. As the modern world continues to pivot towards a digital existence, our reliance on the internet for everyday tasks becomes increasingly pronounced. This heightened reliance necessitates robust defenses, most notably fortified firewall systems, to fortify our networks against these relentless attacks. And herein lies the significance of our work: through the utilization of seven distinct ML models, we've culled that XGBoost emerges as the apex algorithm for this task. And it's not just about identifying the algorithm but also zeroing in on the key features that are indispensable for prediction.

Our findings reveal that a mere five features—namely "Inbound," "Destination Port," "URG Flag Count," "Source Port," and "Avg Bwd Segment Size"—hold the potential to unlock the prevention of DDoS attacks. These features act as beacons, guiding our efforts towards effectively safeguarding our networks. However, we acknowledge that the landscape of cyber threats is perpetually evolving. As we strive to identify common traits of these attacks, the perpetrators themselves are relentless in their quest for innovation. They continually refine their techniques, such as exploiting arbitrary source ports, which serves as a reminder that our battle against these threats is an ongoing one. In this dynamic arena, the onus is on us to keep refining and advancing our machine learning algorithms, equipping them to learn and adapt in tandem with emerging trends and tactics. Ultimately, our journey is fueled by a shared commitment—to protect our networks and digital spaces from the persistent threat of DDoS attacks. Through predictive insights, fortified algorithms, and an unwavering pursuit of knowledge, we are forging a path towards a safer and more secure digital landscape.

FIGURE 8. Firewall and Log Request

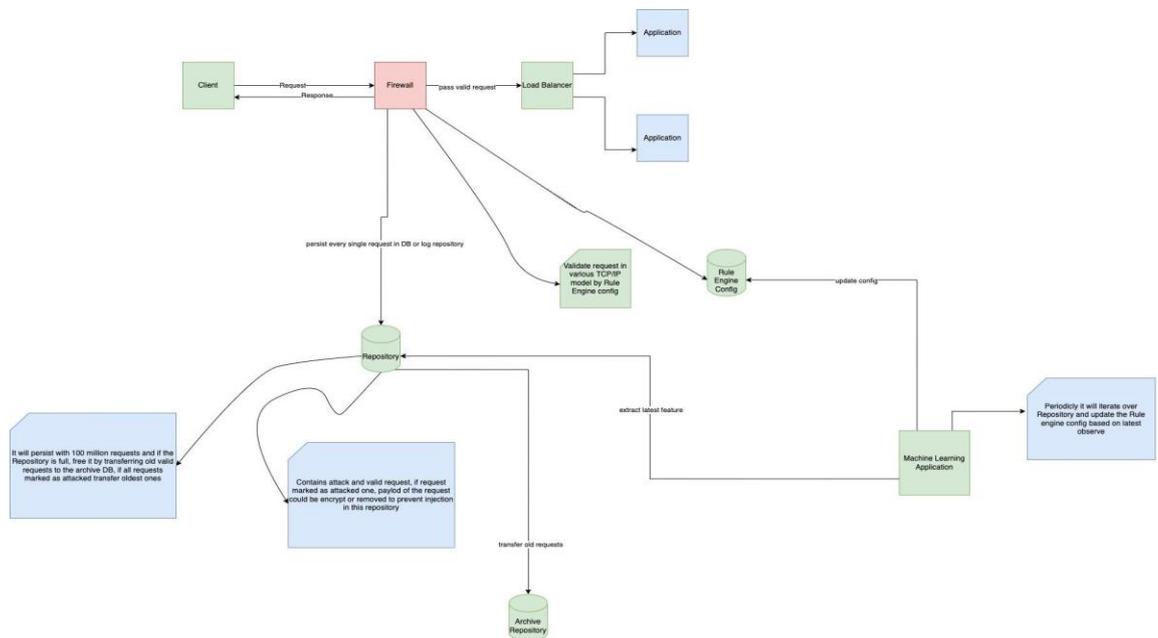

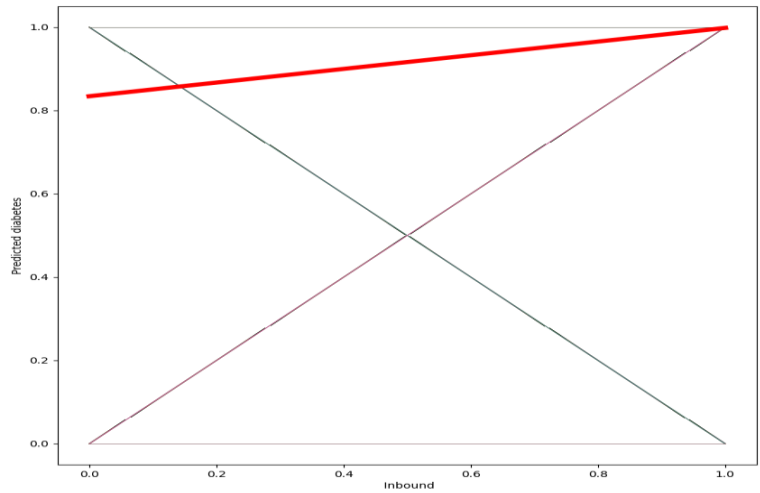

FIGURE 9 Positive relationship between Inbound and Target

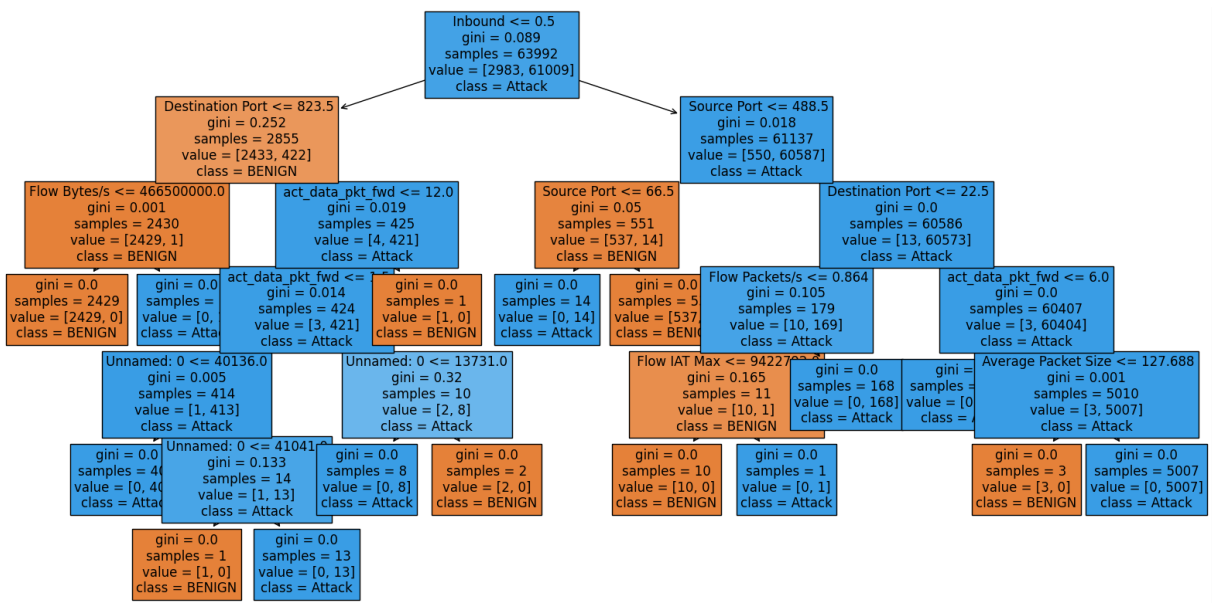

FIGURE 10. Surrogate model (in this case: decision tree)

## ACKNOWLEDGMENT

If the primary author receives an email requesting it, the implementation code for the suggested technique will be made available for study purposes. Additionally, I will post a portion of the code on my GitHub page if anyone is interested in it.